\documentclass[aps,PRB,twocolumn,superscriptaddress,preprintnumbers]{revtex4-2}
\usepackage[T1]{fontenc}
\usepackage{times}
\usepackage{graphicx,color}
\usepackage{amsfonts,amsmath,amssymb,amsbsy}
\usepackage[colorlinks=true, linkcolor=blue, citecolor=blue, urlcolor=blue]{hyperref}
\usepackage{float}

\begin{document}

\title{Braiding and fusion of Majorana fermions in minimal Kitaev spin
liquid \\
on a single hexagon with $5$ qubits}
\author{Motohiko Ezawa}
\affiliation{Department of Applied Physics, University of Tokyo, Hongo 7-3-1, 113-8656,
Japan}

\begin{abstract}
We investigate the minimal Kitaev spin liquid on a single hexagon with three
Ising-type exchange interactions proportional to $K_{x}$, $K_{y}$ and $K_{z}$%
. In the limit $K_{z}=0$, we find 32-fold zero-energy states, leading to 10
free Majorana fermions, and hence, 5 qubits are constructed. These qubits
are protected by particle-hole symmetry even for $K_{z}\neq 0$. Braiding of
these Majorana fermions is possible by temporally controlling a
spin-correlation Hamiltonian. In addition, the fusion is possible by
measuring the spin correlation. By switching on the Heisenberg interaction
together with magnetic field, only one zero-energy state persists, which can
be used as an initialization of qubits. Furthermore, it is shown that $3L+2$
qubits are constructed on the Kitaev spin liquid model on connected $L$
hexagons. All the processes of initialization, operation and readout of
qubits are executable in terms of spin operators.
\end{abstract}

\maketitle

\textbf{Introduction}: A quantum computer is expected to be a most promising
next generation computer\cite{Feynman,DiVi,Nielsen}, which can store $2^{N}$
information in $N$ qubit systems. A topological quantum computation based on
Majorana fermions\cite%
{Brav2,Ivanov,Kitaev01,KitaevTQC,KitaevAP,DasTQC,TQC,EzawaTQC} is
attractive, where $N$ qubits are constructed from $2N$ Majorana fermions.
Majorana fermions are theoretically proposed to emerge in fractional quantum
Hall effects\cite{ReadGreen,ReadBraid,DasTQC,Freedman}, topological
superconductors\cite{Qi,Alicea,Sato,AliceaBraid} and Kitaev spin liquids\cite%
{Kitaev,Jackeli,Motome,Trebst}. The Kitaev topological superconductor model
is the simplest fermionic model that hosts Majorana fermions\cite{Kitaev01}.
Recently, the Minimal Kitaev model consisting of only two sites is realized
in double quantum dots\cite{Dvir,Bordin}. In the view point of quantum
computation, it is desirable to construct a model hosting many Majorana
fermions with the use of smaller number of sites.

\begin{figure}[t]
\centerline{\includegraphics[width=0.48\textwidth]{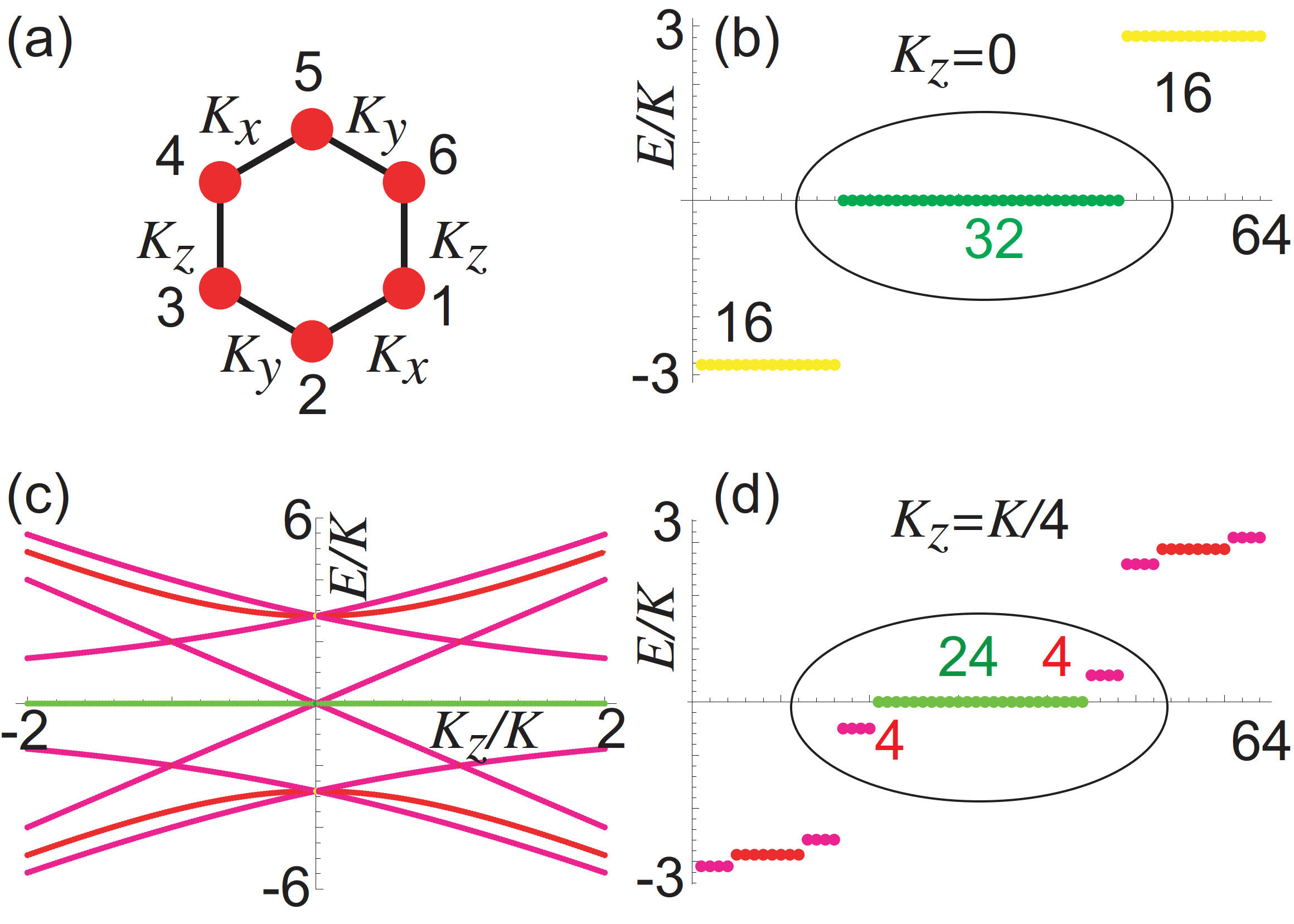}}
\caption{Kitaev spin liquid model on a single hexagon. (a) Illustration of
the Kitaev spin liquid model on a single hexagon, where there are
three-types of Ising interactions. (b) Energy spectrum with $K_{z}=0$. There
are 32-fold degenerate zero-energy states. (c) Energy spectrum as a function
of $K_{z}/K$. There are 24-fold degenerate zero-energy states and 8 nonzero
energy states near the zero energy. (d) Energy spectrum with $K_{z}=K/4$. We
have set $K_{x}=K_{y}=K$. The horizontal axes in (b) and (d) are
eigenindices in the increasing order of the eigenenergy.}
\label{FigUnitCell}
\end{figure}

The Kitaev spin liquid is one of the prominent exactly solvable models on
the honeycomb lattice realizing spin liquid with the emergence of Majorana
zero modes\cite{Kitaev}. It is defined on the honeycomb lattice, where there
are three Ising-type exchange interactions ($\varpropto K_{x},K_{y},K_{z}$)
depending on the directions of bonds as shown in Fig.\ref{FigUnitCell}(a).
By representing the spin operator by a combination of Majorana fermion
operators, the system turns into a free Majorana fermion model on the
honeycomb lattice\cite{Kitaev}. The theoretical proposal on the Kitaev spin
liquid based on perovskite materials\cite{Jackeli} evokes an intensive
researches\cite{Bane,Do,Matsuda}. Experimental signature of Majorana
fermions is observed by measuring a half quantization of thermal conductivity%
\cite{Matsuda,Yokoi}. Brading of Majorana states at vortices in the Kitaev
spin liquid is theoretically proposed\cite{Jang}. In addition to perovskite
material realization, there are several proposals on the realization of the
Kitaev spin liquid model in artificial systems such as qubits\cite%
{Xiao,Bespa}, trapped ions\cite{Ion}, cold atoms\cite{Duan,Sun} and quantum
dots\cite{Cook}. One of the merit of these systems is that it is possible to
construct an extremely small-size system and control model parameters
temporally. The simplest system is a single hexagon. It is an interesting
problem to study whether the Kitaev spin liquid model hosts Majorana
fermions on a single hexagon.

In this paper, we investigate the minimal Kitaev spin liquid on a single
hexagon. When $K_{z}=0$, there are 32=2$^{5}$\ zero-energy states, which
leads to 5 qubits. These qubits are protected by particle-hole symmetry even
for $K_z\neq 0$. Then, we construct braiding operators in terms of spin
operators. A readout process of qubits is performed by the fusion protocol
which measures the local spin correlation. An initialization is executed by
switching on the Heisenberg interaction and magnetic field term, where only
one zero-energy state is present.

\textbf{Minimal Kitaev spin liquid model:} We study the Kitaev spin liquid
model in an artificial system. Especially, we analyze the spin $1/2$ system
on the single hexagon based on the Kitaev model, 
\begin{equation}
\hat{H}_{\text{K}}=\sum_{\alpha =1}^{2}\hat{H}_{\text{K}\alpha }+\hat{H}_{%
\text{K}z},  \label{Basic}
\end{equation}%
with%
\begin{align}
\hat{H}_{\text{K}\alpha } &=-K_{x}\sigma _{3\alpha -2}^{x}\sigma _{3\alpha
-1}^{x}-K_{y}\sigma _{3\alpha -1}^{y}\sigma _{3\alpha }^{y},  \label{HKalpha}
\\
\hat{H}_{\text{K}z} &=-K_{z}\left( \sigma _{3}^{z}\sigma _{4}^{z}+\sigma
_{6}^{z}\sigma _{1}^{z}\right) ,  \label{HKz}
\end{align}%
where $\sigma _{i}^{\gamma }$ is the Pauli matrix at the site $i$ with $%
\gamma =x,y,z$. The Ising-type exchange interaction $K_{\gamma }\sigma
_{i}^{\gamma }\sigma _{j}^{\gamma }$ is anisotropic depending on the
direction of the link $\gamma $ as illustrated in Fig.\ref{FigUnitCell}(a).
It contains $6$ spins, and hence, there are $2^{6}$ states in total.

The Hamiltonian for the Kitaev quantum spin liquid is rewritten in terms of
Majorana fermions by way of the Jordan Wigner transformation\cite%
{Chen1,Feng,Chen2,Nasu1,Nasu2,Hur}. We number the site from 1 to 6 as shown
in Fig.\ref{FigUnitCell}(a). There are relations between the spin operators
and the fermion operators: $\sigma _{j}^{-}=\Omega _{j}c_{j}$, $\sigma
_{j}^{+}=\Omega _{j}c_{j}^{\dagger }$, $\sigma _{j}^{z}=c_{j}^{\dagger
}c_{j}-1/2$ with $\Omega _{j}\equiv \prod\limits_{\ell =1}^{j-1}\exp [i\pi
c_{\ell }^{\dagger }c_{\ell }]$ and $\sigma _{j}^{+}\equiv \frac{1}{2}%
(\sigma _{j}^{x}+i\sigma _{j}^{y})$ and $\sigma _{j}^{-}\equiv \frac{1}{2}%
(\sigma _{j}^{x}-i\sigma _{j}^{y})$. Here, $c_{j}$ and $c_{j}^{\dagger }$
satisfy the anti-commutation relations, $\{c_{i},c_{\ell
}\}=\{c_{i}^{\dagger },c_{\ell }^{\dagger }\}=0$, $\{c_{i},c_{\ell
}^{\dagger }\}=\delta _{i\ell }$. Furthermore, we introduce Majorana
operators as $\gamma _{2j}^{A}=c_{2j}+c_{2j}^{\dagger }$, $\gamma
_{2j}^{B}=(c_{2j}-c_{2j}^{\dagger })/i$, $\gamma
_{2j+1}^{A}=(c_{2j+1}-c_{2j+1}^{\dagger })/i$, $\gamma
_{2j+1}^{B}=c_{2j+1}+c_{2j+1}^{\dagger }$ for $j=1,2,3$. Then, the
Hamiltonians (\ref{HKalpha}) and (\ref{HKz}) read%
\begin{align}
\hat{H}_{\text{K}\alpha }& =-i\left( K_{x}\gamma _{3\alpha -2}^{A}\gamma
_{3\alpha -1}^{A}-K_{y}\gamma _{3\alpha -1}^{A}\gamma _{3\alpha }^{A}\right)
,  \label{HKalpha2} \\
\hat{H}_{\text{K}z}& =-\frac{K_{z}}{4}(\gamma _{3}^{A}\gamma _{3}^{B}\gamma
_{4}^{A}\gamma _{4}^{B}+\gamma _{6}^{A}\gamma _{6}^{B}\gamma _{1}^{A}\gamma
_{1}^{B}),  \label{HKalpha3}
\end{align}%
in the Majorana form.

\textbf{Particle-hole symmetry:} The zero-energy states of the Hamiltonian
with particle-hole symmetry are Majorana fermions\cite%
{Qi,Alicea,Sato,AliceaBraid}. We discuss particle-hole symmetry in the
Kitaev spin liquid. Particle-hole symmetry acts as $P^{-1}\gamma
_{j}^{A}P=\gamma _{j}^{A}$ and $P^{-1}\gamma _{j}^{B}P=\gamma _{j}^{B}$ in
terms of Majorana fermion operators, or $P^{-1}c_{j}P=c_{j}^{\dagger }$ and $%
P^{-1}c_{j}^{\dagger }P=c_{j}$ in terms of fermion operators. In terms of
the spin operators, it acts as%
\begin{align}
P^{-1}\sigma _{j}^{x}P &=\left( -1\right) ^{j-1}\sigma _{j}^{x},\qquad
P^{-1}\sigma _{j}^{y}P=\left( -1\right) ^{j}\sigma _{j}^{y},  \notag \\
P^{-1}\sigma _{j}^{z}P &=-\sigma _{j}^{z},
\end{align}%
where we have used the relation $P^{-1}\Omega _{j}P=\left( -1\right)
^{j-1}\Omega _{j}$.

Under the particle-hole symmetry transformation, the Hamiltonian (\ref{Basic}%
) is mapped to $P^{-1}\hat{H}_{\text{K}}\left( K_{x},K_{y},K_{z}\right) P=-%
\hat{H}_{\text{K}}\left( K_{x},K_{y},-K_{z}\right) $. Hence, the Hamiltonian
has particle-hole symmetry for $K_{z}=0$. We later show that particle-hole
symmetry is present even for $K_{z}\neq 0$ in the present model, as is
consistant with Fig.\ref{FigUnitCell}(d).

\textbf{Minimal Kitaev spin chain models:} We analyze the minimal Kitaev
spin liquid model where $K_{z}$ is much smaller than $K_{x}$ and $K_{y}$. We
first consider the limit $K_{z}=0$ and later include the effect due to $%
K_{z}\neq 0$.

When we set $K_{z}=0$ in Hamiltonian (\ref{Basic}), it is decomposed\cite%
{KitaevAP,Feng,Hur} into two independent Kitaev spin chain models $\hat{H}_{%
\text{K}\alpha }$. There are $2^{3}$\ states because there are 3 spins for
each $\alpha $. By exactly diagonalizing $\hat{H}_{\text{K}\alpha }$ for
each $\alpha $, we find that there are 4-fold degenerate states with $E_{%
\text{K1}}=E_{\text{K2}}=\pm \sqrt{K_{x}^{2}+K_{y}^{2}}$. Note that there
are no zero-energy states in each minimal Kitaev spin chain model.

However, the combined system has $8\times 8=64$ states made of 32
zero-energy states and 32 nonzero-energy states irrespective of $K_{x}$ and $%
K_{y}$. The Hamiltonian (\ref{HKalpha2}) is rewritten in the form%
\begin{equation}
\sum_{\alpha =1}^{2}\hat{H}_{\text{K}\alpha }=\sqrt{K_{x}^{2}+K_{y}^{2}}%
\left( i\gamma _{2}^{A}\bar{\gamma}_{1}^{A}-i\gamma _{5}^{A}\bar{\gamma}%
_{4}^{A}\right) ,  \label{HamilB}
\end{equation}%
where we have defined new Majorana operators%
\begin{equation}
\bar{\gamma}_{1}^{A}=\frac{K_{x}\gamma _{1}^{A}+K_{y}\gamma _{3}^{A}}{\sqrt{%
K_{x}^{2}+K_{y}^{2}}},\qquad \ \bar{\gamma}_{4}^{A}=\frac{K_{x}\gamma
_{4}^{A}+K_{y}\gamma _{6}^{A}}{\sqrt{K_{x}^{2}+K_{y}^{2}}}.
\end{equation}%
Since the Hamiltonian (\ref{HamilB}) does not contain Majorana operators $%
\gamma _{3}^{A}$, $\gamma _{6}^{A}$, $\gamma _{j}^{B}$ with $j=1,2,\cdots ,6$%
, we have $\left[ \hat{H}_{\text{K}},\gamma _{3}^{A}\right] =\left[ \hat{H}_{%
\text{K}},\gamma _{6}^{A}\right] =0$ and $\left[ \hat{H}_{\text{K}},\gamma
_{j}^{B}\right] =0$. Hence, there are 8 free Majorana fermions, from which
we construct 4 fermion operators as $f_{1}^{A}\equiv \left( \gamma
_{3}^{A}-i\gamma _{6}^{A}\right) /2$ and $f_{j}^{B}=\left( \gamma
_{2j-1}^{B}-i\gamma _{2j}^{B}\right) /2$ with $j=1,2,3$. In addition, we
introduce 2 fermion operators $f_{2}^{A}\equiv \left( \bar{\gamma}%
_{1}^{A}-i\gamma _{2}^{A}\right) /2$ and $f_{3}^{A}\equiv \left( \bar{\gamma}%
_{4}^{A}-i\gamma _{5}^{A}\right) /2$. The Hamiltonian (\ref{HamilB}) is
rewritten in terms of these fermion operators as%
\begin{equation}
\sum_{\alpha =1}^{2}\hat{H}_{\text{K}\alpha }=2\sqrt{K_{x}^{2}+K_{y}^{2}}%
\left( \hat{n}_{2}^{A}-\hat{n}_{3}^{A}\right) ,  \label{HamilC}
\end{equation}%
where we have defined the number operators $\hat{n}_{2}^{A}\equiv
f_{2}^{A\dagger }f_{2}^{A}=\left( i\gamma _{2}^{A}\bar{\gamma}%
_{1}^{A}+1\right) /2$ and $\hat{n}_{3}^{A}\equiv f_{3}^{A\dagger
}f_{3}^{A}=\left( i\gamma _{5}^{A}\bar{\gamma}_{4}^{A}+1\right) /2$. In the
similar way, we define $\hat{n}_{j}^{A}=f_{j}^{A\dagger }f_{j}^{A}$ and $%
\hat{n}_{j}^{B}=f_{j}^{B\dagger }f_{j}^{B}$ with $j=1\sim 3$. We consider
the Hilbert space where $\hat{n}_{j}^{A}$ and $\hat{n}_{j}^{B}$ take the
eigenvalues $0$ and $1$. We take $f_{2}^{A}$ to be a free fermion. Then, $%
f_{3}^{A}$\ is determined by the zero-energy condition of the Hamiltonian (%
\ref{HamilC}). As a result, the Kitaev spin liquid model on the single
hexagon contains 10 free Majorana fermions, or 5 qubits defined by $%
\left\vert n_{2}^{A}n_{1}^{A}n_{3}^{B}n_{2}^{B}n_{1}^{B}\right\rangle $.

\textbf{Braiding:} The basic operation on qubits is braiding defined\cite%
{Ivanov} by $\mathcal{B}_{\alpha \beta }=\exp \left[ \frac{\pi }{4}\gamma
_{\beta }\gamma _{\alpha }\right] $. It is generalized\cite{EzawaTQC} to an
arbitrary angle such that $\mathcal{B}_{\alpha \beta }\left( \theta \right)
=\exp \left[ \theta \gamma _{\beta }\gamma _{\alpha }\right] $. The unitary
dynamics under the Hamiltonian $H$ reads $U=\exp \left[ -iHt/\hbar \right] $%
. The generalized brading is executed by setting $-iHt/\hbar =\theta \gamma
_{\beta }\gamma _{\alpha }$. Generalized braiding operators for $B$ Majorana
fermions are rewritten in terms of spin operators as $\exp \left[ \theta
\gamma _{2j-1}^{B}\gamma _{2j}^{B}\right] =\exp \left[ i\theta \sigma
_{2j-1}^{y}\sigma _{2j}^{y}\right] $ for $j=1,2,3$, and $\exp \left[ \theta
\gamma _{2j}^{B}\gamma _{2j+1}^{B}\right] =\exp \left[ i\theta \sigma
_{2j}^{x}\sigma _{2j+1}^{x}\right] $ for $j=1,2$. Generalized braiding
operators for $A$ Majorana fermions are rewritten in terms of spin operators
as%
\begin{align}
& \exp \left[ \theta \bar{\gamma}_{3j-2}^{A}\gamma _{3j-1}^{A}\right]  \notag
\\
& =\exp \left[ \frac{\theta }{\sqrt{K_{x}^{2}+K_{y}^{2}}}\left( K_{x}\gamma
_{3j-2}^{A}\gamma _{3j-1}^{A}-K_{y}\gamma _{3j-1}^{A}\gamma _{3j}^{A}\right) %
\right]  \notag \\
& =\exp \left[ \frac{i\theta }{\sqrt{K_{x}^{2}+K_{y}^{2}}}\left( K_{x}\sigma
_{3j-2}^{x}\sigma _{3j-1}^{x}-K_{y}\sigma _{3j-1}^{y}\sigma _{3j}^{y}\right) %
\right]
\end{align}%
for $j=1,2$. Generalized braiding operators consisting of $A$ and $B$
Majorana fermions are rewritten in terms of spin operators as 
\begin{align}
\exp \left[ \theta \gamma _{2j-1}^{A}\gamma _{2j-1}^{B}\right] & =\exp \left[
-2i\theta \sigma _{2j-1}^{z}\right] , \\
\exp \left[ \theta \gamma _{2j}^{A}\gamma _{2j}^{B}\right] & =\exp \left[
-2i\theta \sigma _{2j}^{z}\right] .
\end{align}%
Hence, it is possible to execute braiding by temporally controlling the spin
Hamiltonian.

\textbf{Fusion:} In order to readout the information of qubits based on
Majorana fermions, the fusion protocol is used, where the fermion number
constructed from Majorana fermions are observed. The fusion is a pair
annihilation process of two Majorana fermions, which results in a single
fermion ($n_{j}=1$) or a vacuum ($n_{j}=0$), and hence, the qubit $n_{j}$
can be readout. The fermion numbers of Majorana fermions are expressed in
terms of spin operators as 
\begin{align}
\hat{n}_{j}^{B}& \equiv f_{j}^{B\dagger }f_{j}^{B}=-i\gamma
_{2j-1}^{B}\gamma _{2j}^{B}/2=-\sigma _{2j-1}^{y}\sigma _{2j}^{y},\quad
j=1,2,3,  \notag \\
\hat{n}_{2}^{A}& \equiv f_{2}^{A\dagger }f_{2}^{A}=-i\bar{\gamma}%
_{1}^{A}\gamma _{2}^{A}=-i\frac{K_{x}\gamma _{1}^{A}+K_{y}\gamma _{3}^{A}}{%
\sqrt{K_{x}^{2}+K_{y}^{2}}}\gamma _{2}^{A}  \notag \\
& =-\frac{K_{x}}{\sqrt{K_{x}^{2}+K_{y}^{2}}}\sigma _{1}^{x}\sigma _{2}^{x}+%
\frac{K_{y}}{\sqrt{K_{x}^{2}+K_{y}^{2}}}\sigma _{2}^{y}\sigma _{3}^{y}.
\end{align}%
Hence, the fusion is executed by measuring the local spin correlation $%
\sigma _{j}^{x}\sigma _{j+1}^{x}$ and $\sigma _{j}^{y}\sigma _{j+1}^{y}$. On
the other hand, it is difficult to readout $\hat{n}_{1}^{A}$ because the
number operator $\hat{n}_{1}^{A}=-i\gamma _{3}^{A}\gamma _{6}^{A}$\ cannot
be represented by a local spin correlation operator.

\textbf{Nonzero }$K_{z}$: We next consider the realistic case with $%
K_{z}\neq 0$. There are conserved quantities\cite%
{Chen1,Feng,Chen2,Nasu1,Nasu2,Hur} known as the $Z_{2}$ gauge fields in the
Hamiltonian (\ref{Basic}). They are real variables $\hat{u}_{34}\equiv
i\gamma _{3}^{B}\gamma _{4}^{B}$ and $\hat{u}_{61}\equiv i\gamma
_{6}^{B}\gamma _{1}^{B}$, satisfying $\left[ \hat{H}_{\text{K}},\hat{u}_{34}%
\right] =\left[ \hat{H}_{\text{K}},\hat{u}_{61}\right] =0$ and $\hat{u}%
_{34}^{2}=\hat{u}_{61}^{2}=1$. The Hilbert space is decomposed into the
subspaces, where $\hat{u}_{34}$ and $\hat{u}_{61}$ take eigenvalues $\pm 1$.
In these subspaces, because the Hamiltonian (\ref{HKalpha3}) becomes in
terms of Majorana fermions as in%
\begin{equation}
\hat{H}_{\text{K}z}=-i\frac{K_{z}}{4}(\hat{u}_{34}\gamma _{3}^{A}\gamma
_{4}^{A}+\hat{u}_{61}\gamma _{6}^{A}\gamma _{1}^{A}),
\end{equation}%
particle-hole symmetry is present. Hence, the Majorana states are
particle-hole symmetry protected even for $K_{z}\neq 0$. Furthermore, it is
possible to diagonalize exactly the Hamiltonian.

We show the energy spectrum as a function of $K_{z}/K$ with $K_{x}=K_{y}=K$
in Fig.\ref{FigUnitCell}(c). There are 24 zero-energy states, 8 states with $%
E=\pm \sqrt{K_{x}^{2}+K_{y}^{2}+K_{z}^{2}}$ and other 24 states as shown in
Fig.\ref{FigUnitCell}(d). The energy spectrum is symmetric with respect to $%
E=0$\ as shown in Fig.\ref{FigUnitCell}(c). For $\left\vert
K_{z}/K\right\vert \ll 1$, 32 states are almost degenerate and it is
possible to use 5 qubits even for $K_{z}\neq 0$.

\textbf{Initialization:} In quantum computation, it is necessary to prepare
one unique quantum state as an initial state. For this purpose, we introduce
the Heisenberg interaction\cite{Cha1,Singh,Cha2} together with the magnetic
field $B_{z}$ along $z$ direction at the initial stage,%
\begin{equation}
\hat{H}_{J}=J\sum_{\left\langle i,j\right\rangle }\sum_{\gamma =x,y,z}\sigma
_{i}^{\gamma }\sigma _{j}^{\gamma }-B_{z}\sum_{j=1}^{6}\sigma _{j}^{z},
\end{equation}%
where we have set $K_{x}=K_{y}=K_{z}\equiv K$. The Hamiltonian $\hat{H}=\hat{%
H}_{\text{K}}+\hat{H}_{J}$ with Eq.(\ref{Basic}) is analytically
diagonalizable for the zero-energy state, and we find it given by 
\begin{align}
\left\vert \psi _{0}\right\rangle \propto &\left\vert 010010\right\rangle
+\left\vert 010101\right\rangle +\left\vert 011001\right\rangle +\left\vert
110001\right\rangle  \notag \\
&-(\left\vert 001100\right\rangle +\left\vert 100100\right\rangle
+\left\vert 101000\right\rangle +\left\vert 101011\right\rangle ).
\end{align}%
The state $\left\vert \psi _{0}\right\rangle $ is used for the
initialization process of the qubits.

\begin{figure}[t]
\centerline{\includegraphics[width=0.48\textwidth]{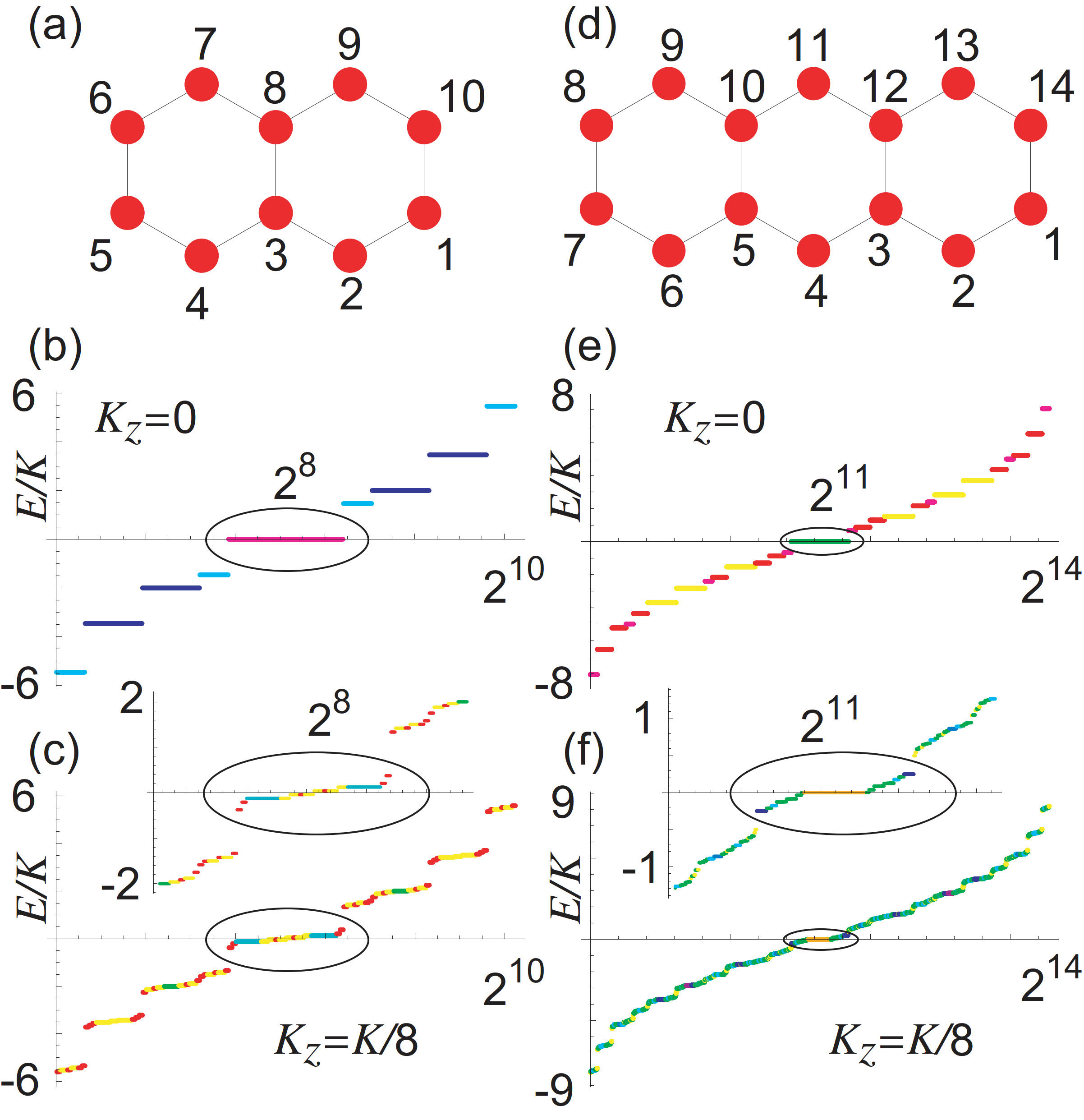}}
\caption{Kitaev spin liquid model on double hexagons and triple hexagons.
(a) Illustration of the Kitaev spin liquid model on double hexagons. Its
energy spectrum with (b) $K_{z}=0$ and (c) $K_{z}=K/8$. (d) Illustration for
the Kitaev spin liquid model on triple hexagons. Its energy spectrum with
(e) $K_{z}=0$ and (f) $K_{z}=K/8$. Insets in (c) and (f) show the enlarged
figures around the zero-energy states. We have set $K_{x}=K_{y}=K$.}
\label{FigLcell}
\end{figure}

\textbf{Kitaev spin liquid on connected hexagons:} It is possible to
generalize the Kitaev spin liquid model on a single hexagon to that on
connected $L$ hexagons, where there are $4L+2$ spins. It is illustrated in
the case of $L=2$\ and $3$\ in Fig.\ref{FigLcell}(a) and (d). The
Hamiltonian reads 
\begin{align}
\hat{H}_{\text{K}}& =-\sum_{j=1}^{L}(K_{x}\sigma _{2j-1}^{x}\sigma
_{2j}^{x}+K_{y}\sigma _{2j}^{y}\sigma _{2j+1}^{y})  \notag \\
& -\sum_{j=1}^{L}(K_{x}\sigma _{2j-1+2L+1}^{x}\sigma
_{2j+2L+1}^{x}+K_{y}\sigma _{2j+2L+1}^{y}\sigma _{2j+1+2L+1}^{y})  \notag \\
& -K_{z}\sum_{j=1}^{L+1}\sigma _{2j-1}^{z}\sigma _{4L+4-2j}^{z}.
\end{align}%
It is rewritten in terms of Majorana fermions as%
\begin{align}
\hat{H}_{\text{K}}=& -i\sum_{j=1}^{L}\left( K_{x}\gamma _{2j-1}^{A}\gamma
_{2j}^{A}-K_{y}\gamma _{2j}^{A}\gamma _{2j+1}^{A}\right)  \notag \\
& -i\sum_{j=1}^{L}(K_{x}\gamma _{2j-1+2L+1}^{A}\gamma _{2j+2L+1}^{A}  \notag
\\
& \qquad \qquad -K_{y}\gamma _{2j+2L+1}^{A}\gamma _{2j+1+2L+1}^{A})  \notag
\\
& -i\sum_{j=1}^{L+1}\frac{K_{z}}{4}u_{2j-1,4L+4-2j}\gamma _{2j-1}^{A}\gamma
_{4L+4-2j}^{A},  \label{HamliMajoL}
\end{align}%
where the $\mathbb{Z}_{2}$ gauge fields are given by%
\begin{equation}
u_{2j-1,4L+4-2j}\equiv i\gamma _{2j-1}^{B}\gamma _{4L+4-2j}^{B},\quad
j=1,\cdots ,L+1.  \label{Z2L}
\end{equation}

We first consider the case $K_{z}=0$, where the system of hexagons is
decomposed into two chains. The system is particle-hole symmetric. The
energy spectrum is shown in Fig.\ref{FigLcell}(b) and (e) for the case of $%
L=2$ and $3$. There are $2^{3L+2}$-fold degenerate\ zero-energy states. It
is understood as follows. There are $2L+1$ free $B$ Majorana fermions
because $\gamma _{j}^{B}$ does not appear in the Hamiltonian of each chain.
On the other hand, there is one free $A$ Majorana fermion according to the
Lieb theorem dictating the number of the zero-energy states in the bipartite
system (\ref{HamliMajoL}). Accordingly, the number of one type of sites is $%
L+1$, while that of the other type of sites is $L$. Hence, the difference is 
$1$ in each chain, implying the presence of one free $A$ Majorana fermion in
Hamiltonian (\ref{HamliMajoL}). Hence, the number of free Majorana fermions
is $2L+2$ in total, which results in the $2^{L+1}$-fold degeneracy in the
energy spectrum. In addition, the diagonalization of the quadratic
Hamiltonian (\ref{HamliMajoL}) for $\gamma _{j}^{A}$ gives different
eigenenergies each other. Hence,\ one Kitaev spin chain model with length $L$
has $2^{L}$ different states with $2^{L+1}$-fold degeneracy. They produces $%
2^{3L+2}=2^{L}\times 2^{L+1}\times 2^{L+1}$-fold degenerate zero energy
states in total. There emerge $2^{3L+2}$ zero-energy states, and hence, $3L+2
$ qubits are constructed.

We next consider the case $K_{z}\neq 0$. The energy spectrum is shown in Fig.%
\ref{FigLcell}(c) and (f) for the case of $L=2$ and $3$, where particle-hole
symmetry holds manifestly. This can be shown with the aid of the $Z_{2}$
gauge fields (\ref{Z2L}) as in the case of the single hexagon. Hence, the
Majorana states are particle-hole symmetry protected even for $K_{z}\neq 0$.

\textbf{Conclusion:} In this paper, we have shown that the Kitaev spin
liquid model on a single hexagon acts as a 5-qubit system. It is possible to
prepare a unique initial state by introducing the Heisenberg interaction
together with magnetic field and to execute braiding by controlling the
local spin correlation Hamiltonian, while qubits are readout with the use of
the fusion protocol by observing local spin correlators.\ In addition, an
arbitrary number of qubits is constructed by using connected hexagons. Our
results are more efficient comparing previous results on the emergence of
Majorana fermions at in the Kitaev spin liquid model.

This work is supported by CREST, JST (Grants No. JPMJCR20T2) and
Grants-in-Aid for Scientific Research from MEXT KAKENHI (Grant No.
23H00171). 


\end{document}